\documentclass[14pt]{extarticle}
\usepackage[bookmarks=true]{hyperref}
\hypersetup{
  pdftitle={Safe Transformative AI via a Windfall Clause},
  pdfsubject={cs.CY, cs.AI},
  pdfauthor={Paolo Bova, Jonas Emanuel M\"uller, Benjamin Harack},
  colorlinks=true,       
  linkcolor=blue,       
  citecolor=blue,       
  filecolor=black,        
  urlcolor=blue,        
  linktoc=page            
}
\usepackage{theapa, rawfonts} 
\usepackage{tikz} 
\usepackage{amsmath} 
\usepackage{IEEEtrantools} 
\usepackage{enumitem} 
\usepackage[capitalise,nameinlink,noabbrev]{cleveref}
\creflabelformat{equation}{#2\textup{#1}#3}
\usepackage[all]{nowidow}
\usepackage{mcdefinitions} 
\usepackage{arxiv} 

\usepackage[T1]{fontenc}    

\usepackage{authblk}

\graphicspath{{./images/}}

\title{Safe Transformative AI via a Windfall Clause}

\DeclareRobustCommand{\orcidicon}{
	\begin{tikzpicture}
	\draw[lime, fill=lime] (0,0) 
	circle [radius=0.16] 
	node[white] {{\fontfamily{qag}\selectfont \tiny ID}};
	\draw[white, fill=white] (-0.0625,0.095) 
	circle [radius=0.007];
	\end{tikzpicture}
	\hspace{-2mm}
}
\foreach \x in {A, ..., Z}{\expandafter\xdef\csname orcid\x\endcsname{\noexpand\href{https://orcid.org/\csname orcidauthor\x\endcsname}
			{\noexpand\orcidicon}}
}
\author[1\thanks{\tt{paolobova@protonmail.com}}]{Paolo Bova}
\author[1\thanks{\tt{jonas.mueller@gmail.com}}]{Jonas Emanuel M\"uller}
\author[1\thanks{\tt{ben.harack@gmail.com}}]{Benjamin Harack\orcidC{}}

\affil[1]{Modeling Cooperation \href{https://www.modelingcooperation.com}{https://www.modelingcooperation.com}}

\renewcommand{\headeright}{Technical Report}
\renewcommand{\undertitle}{Technical Report \#2021-1}
\renewcommand{\shorttitle}{Safe Transformative AI via a Windfall Clause}

\begin{document}
\maketitle

\begin{abstract}
	Society could soon see transformative artificial intelligence (TAI). Models of competition for TAI show firms face strong competitive pressure to deploy TAI systems before they are safe. This paper explores a proposed solution to this problem, a Windfall Clause, where developers commit to donating a significant portion of any eventual extremely large profits to good causes. However, a key challenge for a Windfall Clause is that firms must have reason to join one. Firms must also believe these commitments are credible. We extend a model of TAI competition with a Windfall Clause to show how firms and policymakers can design a Windfall Clause which overcomes these challenges. Encouragingly, firms benefit from joining a Windfall Clause under a wide range of scenarios. We also find that firms join the Windfall Clause more often when the competition is more dangerous. Even when firms learn each other’s capabilities, firms rarely wish to withdraw their support for the Windfall Clause. These three findings strengthen the case for using a Windfall Clause to promote the safe development of TAI.
\end{abstract}


\section{Introduction}\label[section]{sec:introduction}

Progress in artificial intelligence may lead to transformative artificial intelligence (TAI), here defined as artificial intelligence powerful enough to affect ``practically irreversible change that is broad enough to impact most important aspects of life and society'' \cite{gruetzemacher_transformative_2020}. In a future where TAI is developed unsafely, experts in AI safety anticipate risks of grave harm \cite{yudkowsky_intelligence_2013,cave_ai_2018,taddeo_regulate_2018,russell_human_2019,zwetsloot_thinking_2019}.\footnote{We are partial to the definition of TAI discussed at length by \citeA{gruetzemacher_transformative_2020}. Our analysis also holds for their definitions of narrowly or radically TAI as long as those systems promise enormous benefits to the developers of AI alongside risks of grave harm to society. Earlier definitions of TAI include \citeA{muehlhauser_what_2016,dafoe_ai_2018}.} These risks apply not only to the developers of TAI but also to society at large, because of challenges in aligning AI systems to our values\textemdash known altogether as the AI alignment problem \cite{yudkowsky_intelligence_2013}. These challenges are visible in current AI systems \shortcite{amodei_concrete_2016,leike_ai_2017,krakovna_specification_2020}.

Even if researchers find a practical solution to AI alignment, we still have an AI governance problem\textemdash we cannot guarantee that all developers will build TAI safely. The potential for TAI to give a significant competitive advantage to the first mover may motivate firms to build TAI before any of their competitors, which may mean cutting corners on safety.\footnote{Several papers frame this governance problem as an ``AI Race'' \cite{armstrong_racing_2016,naude_race_2020}. However, we recognise the concerns of \citeA{cave_ai_2018} over encouraging such a race narrative. As we often refer to the work of \citeA{armstrong_racing_2016}, mentioning an AI race is unavoidable. Nevertheless, in most places we instead refer to a ``TAI competition''.}

\citeA{armstrong_racing_2016} showed that in a simple model of AI races, it can be rational for each race participant to reduce their own safety investment in order to win, despite this leading to a scenario where the risk of global calamity is high\textemdash a situation in which no one wins. Furthermore, the authors showed that giving agents more knowledge about their position in the race can result in a higher overall disaster risk.

This governance problem is not limited to rational agents in classical game theory. \shortciteA{han_mediating_2021} considered an evolutionary game where firms exhibit social learning. Without policy incentives, firms often ignore safety norms in the hopes of out-competing their rivals \shortcite{han_modelling_2019,han_regulate_2020,han_mediating_2021}.

Furthermore, \citeA{askell_role_2019} identified legal and economic incentives that may encourage AI firms to cut corners on costly safety standards, and drew comparisons to several other industries. As in these other industries, they anticipated that existing competition law is unlikely to regulate firm behaviour on its own.

The danger presented by a competition for TAI has motivated the above researchers and others in AI governance to outline policies which aim to improve the safety of TAI development \shortcite{askell_role_2019,whittlestone_role_2019,han_regulate_2020,clark_regulatory_2020,brundage_toward_2020,tzachor_artificial_2020,vinuesa_role_2020}.

We contribute to this discussion an analysis of the Windfall Clause, a recent policy proposal by \shortciteA{okeefe_windfall_2020}. In a Windfall Clause, firms commit to donating a significant portion of any eventual extremely large profits resulting from TAI to socially good causes.

A Windfall Clause ensures that no single firm will hold too large a share of profits in the event of TAI generating a large windfall. Since any winner must donate a significant share of their windfall profits to socially good causes, a Windfall Clause reduces the risk that firms face when losing the TAI market to competitors \shortcite{ord_moral_2015,jackson_efficiency_2018}. When combined with a model of TAI competition, such as the model by \citeA{armstrong_racing_2016}, we show that this sharing of risk encourages firms to develop TAI more safely.

To realize the Windfall Clause's potential for addressing the AI governance problem, we need to overcome two challenges. First is a coordination problem, firms must want to join the Windfall Clause no matter who else has joined. Second is credibility, firms need to trust that other firms will honour their commitments.

These challenges motivate the following set of questions: how do we design a Windfall Clause which is in every firm's interests to commit to? Will firms still commit when competition is intense and dangerous? Does the availability of information about players' capabilities affect the Windfall Clause's credibility? Our findings for each question are as follows.

Surprisingly, when no clear leader in the competition for TAI has emerged, we can often find a Windfall Clause which eliminates the trade-off between society's expected safety and firms' expected payoffs. Using \citeS{armstrong_racing_2016} model, these \textit{early} Windfall Clauses can in theory guarantee full safety while improving the expected payoffs of all firms. This result helps strengthen the Windfall Clause proposal of \citeA{okeefe_windfall_2020} against concerns that firms would have little incentive to join. As the proposal by \citeA{okeefe_windfall_2020} originally targeted the objective of equity, to avoid the locking-in of harmful values, this ability to promote safe behaviour is surprising on its own.
 
The benefits for firms joining a Windfall Clause are largest when the TAI competition is most dangerous. When competition is more dangerous, the risk-sharing benefits of a Windfall Clause increase, meaning that it is more often rational for a firm to join the Windfall Clause. Precisely when the competition is most dangerous is when firms have the most incentive to cooperate on developing TAI safely. These results for this continuous setting match what is already known about solving discrete social dilemmas via bargaining (such as the prisoner's dilemma): the bargaining range widens as the collective risk from unsafe development grows \cite{harack_technological_2021}.

Even if competition is well underway, the introduction of a \textit{late} Windfall Clause can still make both society and firms better off. When all information about player capabilities is public, we find that the set of late Windfall Clauses which are rational for firms to join is often larger than for early Windfall Clauses. More information often \textit{increases} the incentive to join the Windfall Clause. Even in scenarios where the incentives get weaker with more information, firms are still likely to want to join a late Windfall Clause. This finding demonstrates that pledges firms make in an early Windfall Clause are credible: firms will not want to renege on their pledges upon obtaining more knowledge of the competition.

The next section, \cref{sec:model}, introduces \citeS{armstrong_racing_2016} model of a race for TAI, extending their model to include \citeS{okeefe_windfall_2020} Windfall Clause proposal, and illustrates the effects on competitive dynamics. In \cref{sec:results}, we find conditions for when the Windfall Clause is rational to join and guarantees full safety. We also consider how the set of rational Windfall Clauses changes as the competition becomes more dangerous and as firms learn more about their capabilities. \cref{sec:conclusion} concludes, discussing the implications of these results for designers of a Windfall Clause.

\section{Model}\label[section]{sec:model}

\subsection{Racing to the Precipice}

\subsubsection{Model definition}

\citeA{armstrong_racing_2016} model technological races toward TAI as a static n-player game. In their game, $n$ competitors choose their level of safety, $s_i \in [0, 1]$, which captures how likely they are to avoid disaster. Each firm has a level of capability, $c_i \in [0, \mu]$, which represents how quickly they can develop TAI given their talent, money, and other resources in the absence of safety precautions. The authors assume that implementing the technical standards required for higher safety is costly, and so reduces the speed at which a participant develops TAI. They define a competitor's final score as $\score_i = c_i - s_i$. The winner of the race is the participant who has the highest score and thus develops TAI first.

Note that they consider a uniform distribution of capabilities on the interval $[0, \mu]$. The maximum possible capability, $\mu$, captures the average difficulty of developing TAI relative to the difficulty of achieving full safety. Higher $\mu$ implies that firms likely only need to sacrifice a smaller proportion of their score to reach a given level of safety. 

Once the race ends, there is a risk of disaster if TAI is developed unsafely. \citeA{armstrong_racing_2016} considered this disaster to be existential in nature, and so destroys all the value associated with TAI.\footnote{We note that this model can be applied to a more general notion of disaster. As an example, misaligned TAI might cause widespread and irreparable damage to critical infrastructure \cite{yudkowsky_intelligence_2013,cave_ai_2018}. The economic and human cost could be very large, but society would eventually recover. For \citeS{armstrong_racing_2016} model, it is still important to note that the payoffs to all firms in the event of disaster is set to $0$, which implies that they assume the disaster affects all the firms equally. This situation might occur if the disaster led to a complete ban on TAI technologies.} The safety of the winner, $s_{\textrm{top}}$, tells us the likelihood of disaster. At the extremes, $s_{\textrm{top}} = 0$ guarantees a disaster and $s_{\textrm{top}}=1$ makes disaster impossible. 

Firms may prefer other firms winning the TAI competition over a disaster\textemdash perhaps the first mover advantage is not especially strong or perhaps a disaster threatens the wider economy that other firms care for.  \citeA{armstrong_racing_2016} captured these preferences with the concept of \textit{enmity} between competitors. Enmity, $e \in [0, 1]$, tells us how a firm compares the utility of a disaster with the utility of another firm winning the race. For example, $e=0.4$ means that another firm winning is regarded as 40\% as bad as disaster.

We can now define the payoffs to firms as follows:

\begin{equation}
    u_i(s) = \begin{cases}
                s_i  & \score_i = \score_{\textrm{top}}\\
                (1-e) s_{\textrm{top}} & \score_i < \score_{\textrm{top}}
             \end{cases}
    \label{eq: payoffs}
\end{equation}

If a firm has the highest score, they win the race. The winner avoids disaster with chance $s_i = s_{\textrm{top}}$, and only then enjoys the benefits of winning the race. The utility from these benefits are normalised to $1$, meaning a payoff of $s_{\textrm{top}}$. Losers still see a utility of $(1-e) s_{\textrm{top}}$ since they may prefer another firm winning to disaster.

Given these model primitives, firms have incentive to cut their safety to increase their chance of winning the race, but doing so encourages their competitors to do the same. Hence, we have a race to the bottom on safety.

The objective of firms is to maximize their expected payoffs subject to the available information. \citeA{armstrong_racing_2016} derived optimal strategies for three information conditions for agents: 1) no knowledge, 2) knowledge about their own capability, and 3) knowledge about everyone's capabilities. The authors solved for a Nash equilibrium in information conditions 1 and 3, and a Bayesian Nash equilibrium in information condition 2. Therefore, competitors play best responses to each other's strategies given the available information. For ease of exposition, we will only consider their full information condition. Let $\Delta$ be the difference between the capability ($c$) of the top firm and the second ranked firm. \cref{eq:full_rttp_s} shows the winner's optimal safety under the full information condition.

    \begin{equation}
        s_{\textrm{top}} = \begin{cases}
         \frac{\Delta}{e} & \Delta < e \\
         1 & \Delta \geq e
        \end{cases}
        \label{eq:full_rttp_s}
    \end{equation}

\citeA{armstrong_racing_2016} used these optimal strategies to determine the average risk of disaster over the distribution of capabilities. The disaster risk for \citeS{armstrong_racing_2016} full information model is: 
    \begin{equation}
        \text{Disaster Risk} = \begin{cases}
         1 - \frac{\mu}{e  (n + 1)} & \mu < e \\
         1 - \frac{\mu}{e  (n + 1)}  + \frac{(\mu - e)^{n+1}}{e  (n + 1)\mu^n} & \mu \geq e \\
        \end{cases}
        \label{eq: full info AI Risk}
    \end{equation}

Using \cref{eq: full info AI Risk} we can compute the expected safety:

\begin{equation}
    E[\textrm{Safety}] = 1 - \textrm{Disaster Risk}
    \label{eq: expected safety}
\end{equation}

Note that after we average over capabilities, each firm has an equal chance of winning the race. Thus, expected payoffs are:

\begin{equation}
    E[\textrm{Payoff}] = \frac{1}{n} E[\textrm{Safety}] + \frac{n-1}{n} (1-e) E[\textrm{Safety}] 
    \label{eq: expected payoffs}
\end{equation}

\subsubsection{Suitability for Windfall Clause analysis}

\citeS{armstrong_racing_2016} model is sufficient to capture the impact of the Windfall Clause. We inherit the assumptions of their model and game-theoretic solution method, but their model's simplicity suggests that there may exist a wide family of similar models for which our main conclusions still hold.

The broader literature on AI races displays a range of complex behaviours. \citeA{han_regulate_2020} introduced a model of AI races which uses evolutionary game theory. Their model is also amenable to a Windfall Clause analysis. \citeA{naude_race_2020} have a model of AI races as a bidding contest, although they do not discuss safety. As these models and others show, players can make many other decisions which influence race dynamics: they advertise, lobby, collaborate, disclose information, monitor, reward, or punish other players in related domains \cite{bar_defensive_2006,fernandes_norms_2019,han_modelling_2019,han_regulate_2020,han_mediating_2021,naude_race_2020,li_multilateral_2020,lacroix_tragedy_2021}.

Nevertheless, \citeS{armstrong_racing_2016} model includes all the relevant features for raising awareness of the efficiency gains from safely developing TAI. When introducing new behaviours helps us encourage better outcomes for society, without worsening others, we should consider adding them to our framing of the game. Our motivation for introducing the decision to join a Windfall Clause is to highlight such a behaviour. 

\subsection{Windfall Clause}
 
\subsubsection{Definition}

\citeA{okeefe_windfall_2020} have proposed a Windfall Clause, where any discoverer of TAI breakthroughs commits to donate a significant percentage of any eventual extremely large profits to good causes. 

We now propose a simple model of the Windfall Clause. This model defines how much of the prize firms should donate and which causes are acceptable candidates for donation. We therefore introduce:

\begin{itemize}
\item \textit{Pledged windfall}, $\tau \in [0,1]$, the share of the prize donated by the winning firm. We assume the prize is divisible and fungible, and without loss of generality will continue to refer to the prize as profits. The prize from developing TAI could also refer to market share or economic power.
\item \textit{Donation enmity}, $e_{wc} \in [0,1]$. Similar to enmity between firms, donation enmity tells  us  how  a  firm  compares  the  utility  from donations via the Windfall Clause with the utility from disaster. For example, $e_{wc}=0.4$ means that firms regard the donation going to socially good causes as 40\% as bad as a disaster. Donation enmity therefore depends on the mechanism which allocates windfall profits to causes. If competitors can only donate to highly effective causes, they will tend to have a different donation enmity than if all competitors had equal say in where the donations should go.
\end{itemize}

Our results do not require that the Windfall Clause triggers for certain. Firms may instead see TAI as a remote possibility. As long as that remote possibility involves the same trade-off in safety and capability as \citeS{armstrong_racing_2016} model, then all our findings hold. In particular, this means that we assume firms cannot experience windfall profits without risking disaster. 

\citeA{okeefe_windfall_2020} introduced this framing of windfall being a remote possibility for AI firms. Since windfall is unlikely, firms joining a Windfall Clause commit little of their future expected revenues away. Under this framing, even pledging the entire possible windfall, $\tau=1$, might seem reasonable to firms, especially if it were to guarantee full safety compliance from all firms. Firms can commit to a Windfall Clause now and that Windfall Clause will only become relevant to their decision making once firms believe that developing TAI is plausible. Thus, whenever we find below that firms are better off joining a Windfall Clause, they are better off no matter how likely they believe TAI to be. Without loss of generality, we assume that the winner of the TAI competition triggers the Windfall Clause for certain.

A Windfall Clause with pledged windfall, $\tau$, and donation enmity, $e_{wc}$, changes the payoffs in \cref{eq: payoffs} to:

\begin{equation}
    u_i(s) = \begin{cases}
                ((1-\tau) + \tau (1-e_{wc})) s_i  & \score_i = \score_{\textrm{top}}\\
                ((1-\tau) (1 - e) + \tau (1-e_{wc})) s_{\textrm{top}} & \score_i < \score_{\textrm{top}}
    \end{cases}
\end{equation}

The winner keeps fraction $1 - \tau$ of the profits and gives fraction $\tau$ of the profits to causes with donation enmity $e_{wc}$. The losers' payoffs reflect how their enmity towards the winner may differ from their enmity towards the winner's donations.

Factorizing the above payoffs lets us simplify to:

\begin{equation}
    u_i(s) = \begin{cases}
                \lambda_{\pi} s_i  & \score_i = \score_{top}\\
                \lambda_{\pi} (1 - \lambda_e e) s_{\textrm{top}} & \score_i < \score_{top} 
    \end{cases}
    \label{eq: payoffs_windfall}
\end{equation}
 
\begin{equation}
    \lambda_{\pi} = (1-\tau) + \tau (1-e_{wc}) = 1 - \tau e_{wc}
\end{equation}

\begin{equation}
    \lambda_e = \frac{1 - \tau}{(1-\tau) + \tau (1-e_{wc})} = \frac{1 - \tau}{ 1 - \tau e_{wc}}
\end{equation}

The Windfall Clause specified above has two consequences for \citeS{armstrong_racing_2016} model. First, the pledged windfall means that firms sacrifice some profits to causes that are only partially aligned with their values (and the donation enmity is the strength of misalignment). Notice that $\lambda_{\pi}$ represents the value of profits under the pledged windfall. Winners and losers both share the common factor $\lambda_{\pi}$, so it does not affect the optimal behaviour of competitors in the race model (\cref{eq:full_rttp_s_windfall}).

Second, the Windfall Clause has a risk sharing benefit which can reduce the danger from TAI competition. Since all firms commit to donating some profits to socially good causes if they experience windfall, firms effectively compete for a smaller part of the market. As the gap between winner and loser shrinks, the Windfall Clause moves us away from a winner-take-all scenario. This reduction in competitive pressure shrinks the effective enmity by a factor $\lambda_e$. For the purposes of decision making, the only change to the game due to the Windfall Clause is that enmity has shrunk. We can express the winner's optimal safety under the Windfall Clause as:

\begin{equation}
        s = \begin{cases}
         \frac{\Delta}{\lambda_e e} & \Delta < \lambda_e e \\
         1 & \Delta \geq \lambda_e e
        \end{cases}
        \label{eq:full_rttp_s_windfall}
    \end{equation}
    
As enmity falls by factor $\lambda_e$, firms are willing to invest more in safety, reducing the risk of disaster.

Note that the expected payoff for each firm without a Windfall Clause, $E[\textrm{Payoff}]$ (\cref{eq: expected payoffs}), is a function of $e,n,\mu$. We will let $E\Pi(e,n,\mu)$ denote the $E[\textrm{Payoff}]$ without a Windfall Clause. The effects of the Windfall Clause discussed above lead to a new expected payoff of $\lambda_{\pi} E\Pi(\lambda_e e,n,\mu)$.

\subsubsection{Implementation}

Now that we have specified how the Windfall Clause functions, we must also consider how to implement it. Any implementation of the Windfall Clause should explain how it overcomes two challenges. The first challenge is coordinating firms to join a Windfall Clause whose benefits are not exclusive to those who join. The second challenge is building the credibility of the Windfall Clause so that firms can trust the winner to honour their commitments.

The literature on assurance contracts gives us one way to design a Windfall Clause which coordinates firms to join. In line with the work of \citeA{bagnoli_provision_1989} or \citeA{tabarrok_private_1998}, a firm only triggers the Windfall Clause if they reach windfall and all competitors have signed up to the Windfall Clause. This way, the payoffs only change once everyone has joined the Windfall Clause. We can then say that joining the Windfall Clause is rational as long as everyone joining is better than no one joining.

The challenge of building credibility manifests in two ways. The largest concern is whether firms believe other firms will honour their commitments. \citeA{okeefe_windfall_2020} discussed this challenge further: competitors who experience windfall may even gain the power to disrupt legal proceedings, so may renege on their Windfall Clause commitments. As our results will show, firms have a lot to gain from establishing a Windfall Clause which promotes safety. Therefore, firms have a powerful incentive to find credible ways to commit to a Windfall Clause.

However, there is a second concern for the credibility of the Windfall Clause. Will firms have incentive to back-out of the Windfall Clause once they learn more information about everyone's capabilities? To answer this question, we compare two scenarios for designing a Windfall Clause. First is the \textit{early} scenario where firms join a Windfall Clause before any clear leader emerges among them. The competition begins sometime after firms choose their Windfall Clause, at which time the information available to firms may differ.\footnote{We focus on a competition under full information for the sake of concision. Except where stated the results for the other information conditions are the same.} Second, we have a \textit{late} scenario where firms decide on a Windfall Clause after all firms know each other's capabilities.\footnote{These scenarios are analogous to choosing a Windfall Clause under no information and full information, as in the work of \citeA{armstrong_racing_2016}. We do not examine a Windfall Clause chosen under private information. Such a proposal is more difficult to analyse since any pledges made by firms also act as a signal of their hidden capabilities. We will consider further research into designing mechanisms which encourage firms to join a Windfall Clause even if it means revealing their hidden capabilities.} By comparing these two scenarios, we show that firms rarely have a strong incentive to back out of an early Windfall Clause once they enter the late scenario.

\section{Results}\label[section]{sec:results}

\subsection{Firms typically have incentives to join a Windfall Clause which promotes full safety}

We find that, for a surprisingly large range of donation enmity values, firms achieve a higher expected payoff after joining a Windfall Clause, even as the pledged windfall approaches $1$. Since a higher pledged windfall implies that the Windfall Clause has a higher impact on safety, this finding is very encouraging for those who wish to use the Windfall Clause to reduce the risk of disaster from TAI.

\begin{figure}[ht]
    \centering
    \includegraphics[width=0.6\textwidth]{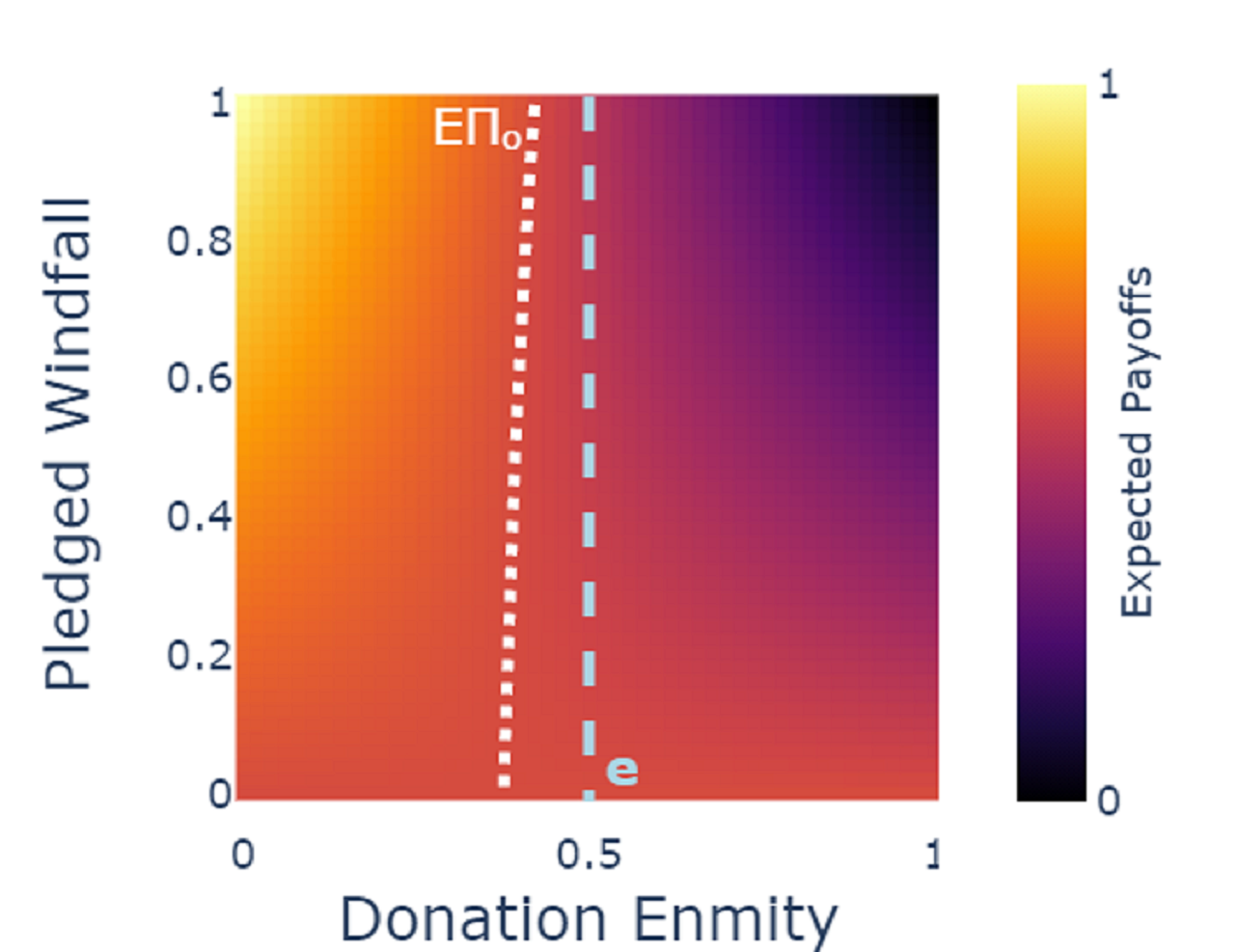}
    \caption{\small Expected payoffs to firms under different Windfall Clauses. The curve $E\Pi_o$ shows where a Windfall Clause provides exactly the same expected payoffs as having no Windfall Clause (pledged windfall $ = 0$). All Windfall Clauses to the left of the $E\Pi_o$ curve raise the expected payoffs above their expected payoffs without a Windfall Clause. These Windfall Clauses are rational for firms to join. Model parameters are: $n=2, \mu=2, e=0.5$.}
    \label{fig: expected payoffs default}
\end{figure}

\cref{fig: expected payoffs default} shows the impact of the Windfall Clause on expected payoffs, defined in \cref{eq: expected payoffs}, as we vary the pledged windfall and the donation enmity. We first show results for a typical competition with parameter values, $n=2, \mu=2, e=0.5$, since such a competition is neither especially safe nor especially dangerous. \cref{sec:result2} discusses what happens as we change these values, confirming that firms often have powerful incentives to join a Windfall Clause.

Firms receive the highest expected payoffs when the donation enmity is low and pledged windfall is high. Expected payoffs are low if donation enmity is high and the pledged windfall is high. These patterns are unsurprising. If firms perceive the value of their donations via the Windfall Clause to be almost as valuable as keeping their profits, then the Windfall Clause is a small price for improving safety. If firms perceive their donations via the Windfall Clause to be almost as bad as disaster, then the Windfall Clause has a much smaller risk-sharing benefit and thus fails to improve safety or firm payoffs. 

However, notice that an extensive region of \cref{fig: expected payoffs default} (the region to the left of indifference curve $E\Pi_0$) finds an expected payoff greater than the expected payoff with no Windfall Clause. For a donation enmity, $e_{wc} \leq 0.4$, we can say that it is rational for firms to join a Windfall Clause with pledged windfall, $\tau = 1$. 

One may wonder why the strongest Windfall Clause is most likely to make firms better off when intuition might suggest the opposite. The reason is that, at least in the full information model of \citeA{armstrong_racing_2016}, firm safety in equilibrium is very responsive to the changes in enmity induced by the Windfall Clause. Consequently, when donation enmity is not too high, the safety benefits of increasing the pledged windfall always outweigh any foregone profits \cite{armstrong_racing_2016}.

From now on, we will name the largest donation enmity for which firms will rationally join a Windfall Clause with pledged windfall high enough to guarantee full safety as the donation enmity limit, $e_{wc}^*$. In the case of an early windfall clause, this definition implies a pledged windfall of $\tau = 1$. Notice that in \cref{fig: expected payoffs default} the donation enmity limit $e_{wc}^*=0.4$ is not too much lower than the value of enmity towards other firms, $e=0.5$. Thus, for many plausible values of the donation enmity, the interests of firms align with society's concern for safety. We will show in the next section that the donation enmity limit remains high as we explore the rest of the parameter space.

\subsection{Firms are more likely to join a Windfall Clause as the competition becomes more dangerous}\label{sec:result2}

For the Windfall Clause to be a robust policy for improving safety, we need firms to want to join it even when the danger of the competition is high. We consider four ways that the competition becomes more dangerous: higher enmity ($e$) between the agents, a larger number of competitors ($n$), a lower max capability ($\mu$), and increased availability of information.\footnote{This finding considers the information firms have during the competition, after choosing the Windfall Clause. In this section (\cref{sec:result2}), we assume firms choose a Windfall Clause with no information about firm capabilities.} Each time, we find that as the competition becomes more dangerous, the donation enmity limit rises. Therefore, when the competition is more dangerous, we are more likely to find Windfall Clauses which firms will rationally join.

We now derive an expression for the donation enmity limit. Recall that, in the current setting, the donation enmity limit is the largest donation enmity for which firms will rationally join a Windfall Clause with pledged windfall ($\tau$) equal to $1$. We first note that we can describe the set of rational windfall clauses with pledged windfall, $\tau$, for a given donation enmity, $e_{wc}$ as: 

\begin{equation}
    \{(e_{wc}, \tau): \lambda_{\pi} E\Pi(\lambda_e e, n, \mu) \geq E\Pi(e, n, \mu)\}
    \label{eq: set of early rational windfall clauses}
\end{equation}

For the donation enmity limit, we are interested in the boundary of this set. On this boundary, the expected payoffs from joining the Windfall Clause will be equal to the expected payoffs without a Windfall Clause:
\begin{equation}
    \lambda_{\pi} E\Pi(\lambda_e e, n, \mu) = E\Pi(e, n, \mu)
    \label{eq: expected payoff indifference windfall clause}
\end{equation}

When $\tau=1$: $\lambda_{\pi} = 1-e_{wc}^*$ and $\lambda_e = 0$. Moreover, $E\Pi(0,n,\mu) = 1$. Rearranging for the donation enmity limit gives us:

\begin{equation}
    e_{wc}^* = 1 - E\Pi(e, n, \mu)
    \label{eq: donation enmity limit}
\end{equation}

\begin{figure}[ht]
    \centering
    \includegraphics[width=\textwidth]{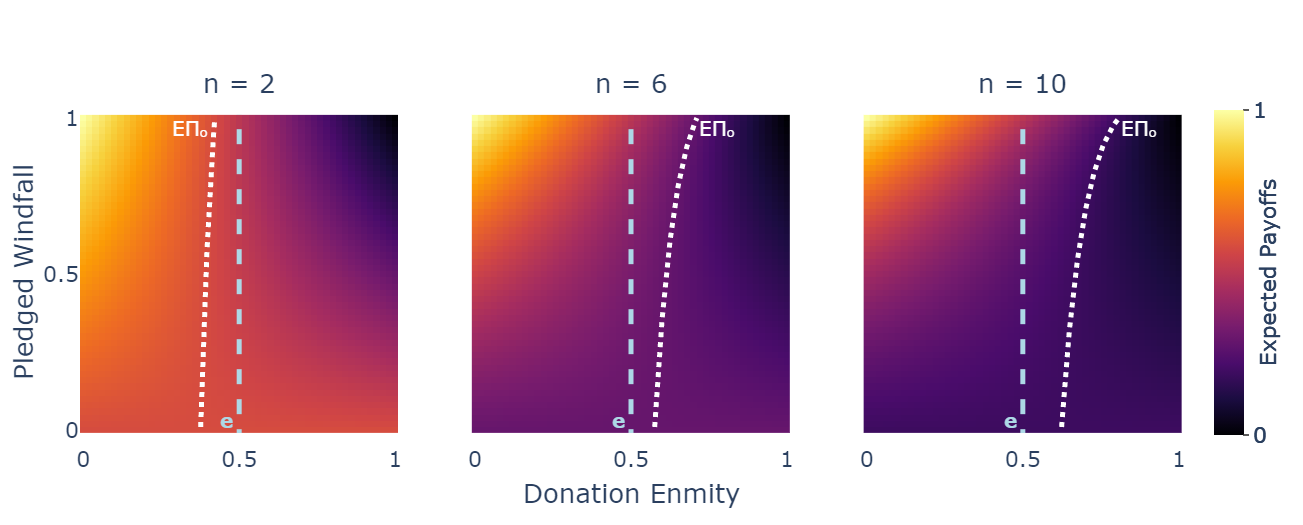}
    \caption{\small Expected payoffs to firms under different Windfall Clauses as we vary the number of competitors, $n$. The curve $E\Pi_o$ shows where a Windfall Clause yields the same expected payoff as having no Windfall Clause (pledged windfall $ = 0$). Windfall Clauses to the left of the curve make firms better off, so we say they are rational for firms to join. Increasing $n$ makes the competition more dangerous, which pushes $E\Pi_o$ to the right. For higher $n$, $E\Pi_o$ is to the right of $e$, meaning it is even rational for firms to join a Windfall Clause with \textit{higher} donation enmity than their enmity towards other firms.}
    \label{fig: expected payoffs vary n}
\end{figure}

We first consider a larger number of competitors, $n$. \cref{fig: expected payoffs vary n} shows the expected payoffs from \cref{fig: expected payoffs default} as we vary $n$. Notice that without a Windfall Clause, increasing $n$ reduces the expected payoffs. As $n$ grows, the gap between the top and second top capabilities is more often smaller. This competition provokes firms to invest less in safety, so the expected safety falls. Higher $n$ also reduces each firm's chance of winning the competition, so expected payoffs also fall.

As one might expect, since firms are receiving a low payoff without a Windfall Clause, there is a larger region of Windfall Clauses which can make firms better off. In \cref{fig: expected payoffs vary n}, the curve $E\Pi_o$, representing indifference between a Windfall Clause or no Windfall Clause (\cref{eq: expected payoff indifference windfall clause}) shifts to the right as $n$ increases.

\begin{figure}[ht]
    \includegraphics[width=\textwidth]{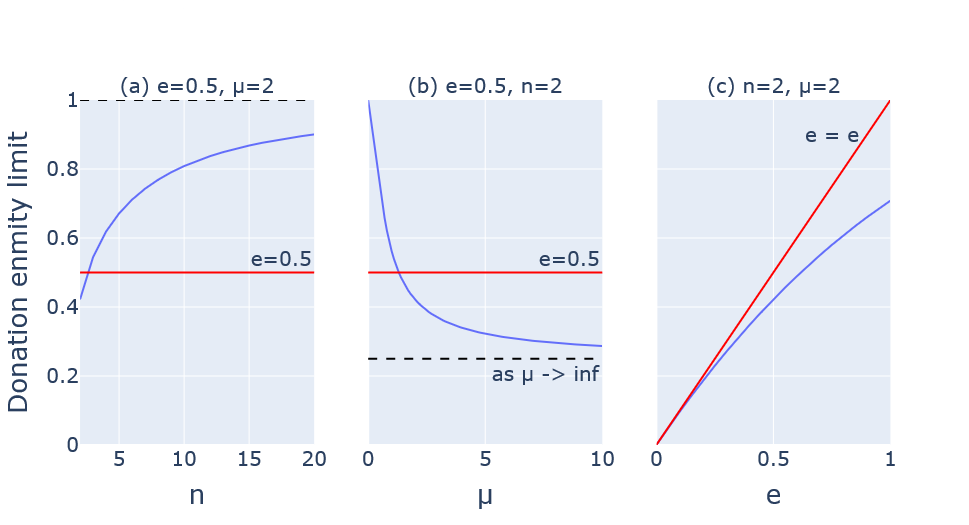}
    \caption{\small The donation enmity limit as we vary different parameters. As the competition becomes more dangerous (higher $n$, lower $\mu$, higher $e$),  firms will join a Windfall Clause for higher values of donation enmity, raising the donation enmity limit (blue line). For high $n$ or low $\mu$, firms will rationally join a Windfall Clause even when their donation enmity is \textit{higher} than their enmity (red line).\\
    (a) varies the number of competitors, $n$. $e=0.5, \mu=2$ \\
    (b) varies the max capability, $\mu$. $e=0.5, n=2$ \\
    (c) varies the enmity towards other competitors, $e$. $n=2, \mu=2$
}
    \label{fig: donation enmity limit vary all}
\end{figure}

Moreover, the donation enmity limit $e_{wc}^*$ increases too. In fact, $e_{wc}^*$ increases above $e$. With only a small increase in the number of agents, firms will sometimes want to join a Windfall Clause even when they would prefer their competitors winning over donating their winnings via the Windfall Clause. As we can see in \cref{fig: expected payoffs vary n}, the gap between the donation enmity limit and enmity towards other firms increases with higher $n$. In \cref{eq: donation enmity limit}, this tells us that the expected payoff as $n$ rises becomes tiny. Intuitively, increasing $n$ leads to a competition so dangerous that the safety benefits of a Windfall Clause with high pledged windfall become enormous.

We see a somewhat similar pattern when changing the other parameters. As the difficulty of achieving TAI relative to safety, $\mu$, decreases, the capabilities of firms will be closer together. When firms are closer in terms of capabilities, they experience more competitive pressure to cut their safety spending. As $\mu$ falls in \cref{fig: donation enmity limit vary all}, we see that the donation enmity limit rises. For sufficiently low $\mu$, once again, the donation enmity limit surpasses enmity. This means that for sufficiently low $\mu$ firms will want to join a Windfall Clause even when they would prefer their competitors winning over donating the winnings via the Windfall Clause.\footnote{We only consider values for $\mu$ in $[0, 10]$. One can show that when $n=2$ and $\mu=10$, even without a Windfall Clause, the winning firm is very likely ($>95\%$ chance) to guarantee full safety. This is also the reason why the donation enmity limit approaches an asymptote. For high enough $\mu$, the expected safety approaches $1$. From \cref{eq: donation enmity limit} we can see that the donation enmity limit will approach $1 - \frac{n(1-e) + e}{n} = 0.25$ in \cref{fig: donation enmity limit vary all}.}

Higher enmity, $e$, moves us closer to a winner-take-all scenario, so firms reduce their safety spending. As expected, the donation enmity limit is increasing in $e$, see \cref{fig: donation enmity limit vary all}. However, this time, notice that the donation enmity limit does not surpass $e$ (in this case $n=2$), and the difference between the two grows for higher enmity.

As for the role of information, recall that \citeA{armstrong_racing_2016} found in their model that the danger of the race often increased as firms had more information about their capabilities, a result backed up by the literature on rivalry \shortcite{bar_defensive_2006,chen_does_2017,zucchini_competitive_2019,hutzschenreuter_competitors_2021}. In line with our above findings, if firms anticipate they will have public information rather than private or no information, then they believe that on average that the race is more dangerous. Firms have more to gain from a Windfall Clause, so their donation enmity limits are higher too. 

Specifically, \citeA{armstrong_racing_2016} found that the ``no information'' condition is the safest and that in many cases private information is safer than public. However, public information can be safer than private when enmity is low enough. Our findings mirror their results. No information leads to the lowest donation enmity limits, private information has higher donation enmity limits than no information but these are only higher than for full information when enmity is low enough.

\begin{figure}[ht]
    \includegraphics[width=\textwidth]{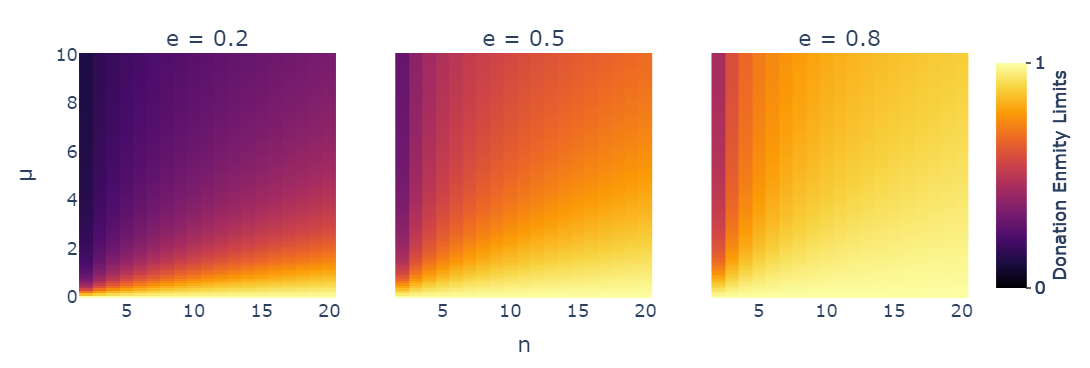}
    \caption{\small Donation enmity limit as we vary the number of competitors, $n$, and max capability, $\mu$, for different values of $e$. When $e$ is low, the donation enmity limit is often low too. When $e$ is high, the donation enmity limit is often very high, especially for high $n$.}
    \label{fig: donation enmity limit robust in e}
\end{figure}

In \cref{fig: donation enmity limit robust in e}, we confirm that these relationships from \cref{fig: donation enmity limit vary all} hold throughout the parameter space. Especially for high $e$, but also for high $n$ and low $\mu$ even when $e$ is low, the donation enmity limit becomes very large, often surpassing enmity, and at times approaching $1$. In the most dangerous scenarios, firms only have to prefer the Windfall Clause slightly over disaster to want to join a Windfall Clause which fully averts disaster. 

Despite these findings, we do not recommend that policymakers actively increase the danger of the competition. Politically, convincing firms to join the strongest Windfall Clause may prove very demanding. If for any reason the proposal breaks down, or even if the Windfall Clause is weaker than hoped for, measures to increase the danger of the competition may leave society worse off than before. Furthermore, policymakers who take steps which actively increase danger could find it hard to build the credibility to convince firms that a Windfall Clause is in their best interest. We advise that this finding is mainly used to advocate for a Windfall Clause: even in the most dangerous scenarios, firms have every reason to join a strong Windfall Clause.

\subsection{Common knowledge of capabilities further incentivizes firms to join a Windfall Clause}

Until now, we have assumed firms do not know their capabilities when choosing a Windfall Clause. They learn those capabilities only once the competition has begun. However, if we expect firms to commit to a Windfall Clause voluntarily, then we might wonder whether firms might try to leave the Windfall Clause once they know their capabilities. 

To answer this important question, we distinguish between an \textit{early} Windfall Clause where firms lack knowledge of capabilities when choosing a Windfall Clause (as discussed above) and a \textit{late} Windfall Clause where firms choose a Windfall Clause with full knowledge of everyone's capabilities.\footnote{The labels ``late'' and ``early'' are heuristic. We anticipate that information about firm capabilities is more likely to become available as time progresses. From a policy perspective, this distinction between an early and late Windfall Clause seems applicable. Note that \citeA{okeefe_windfall_2020} referred to an ex-ante Windfall Clause, which is equivalent to our early Windfall Clause. We avoid using an ``ex-ante'' versus ``ex-post'' distinction to prevent confusion with the other uses of these terms in game theory.} Fortunately, we find that firms often have a stronger incentive to join a late Windfall Clause than an early one. Therefore, firms are unlikely to face future incentives to back out of an early Windfall Clause, dramatically increasing its credibility. 

To gain an intuition for why firms are more likely to join the late Windfall Clause, we explain how to find the set of late Windfall Clauses which are rational for firms to join. It is easy to show from the payoffs in \cref{eq: payoffs} that if a winner finds the Windfall Clause rational, then so will the losers. The winner finds the Windfall Clause rational if their utilities from \cref{eq: payoffs} are higher under the Windfall Clause, \cref{eq: payoffs_windfall}. Recall that the Windfall Clause reduces enmity by factor $\lambda_e$ and firm utilities by factor $\lambda_{\pi}$. Denote the winner's safety without the Windfall Clause, \cref{eq:full_rttp_s}, as $s_{\textrm{top}}$ and with the Windfall Clause, \cref{eq:full_rttp_s_windfall}, as $s$. We have three cases:
\begin{itemize}
    \item The leader is guaranteed safety without a Windfall Clause, $s_{\textrm{top}} = 1$: the leader is weakly worse off under a Windfall Clause whenever $\lambda_{\pi} \leq s_{\textrm{top}} = 1$.
    \item Windfall Clause pushes leader to increase safety to $s = 1$ from $s_{\textrm{top}} < 1$: the leader is better off if $\lambda_{\pi} \geq s_{\textrm{top}}$.
    \item Windfall Clause pushes leader to increase safety to $s= \frac{\Delta}{\lambda_e e}  =\frac{s_{\textrm{top}}}{\lambda_e} < 1$ from $s_{\textrm{top}} < 1$: the leader is better off if $ \lambda_{\pi} s = s_{\textrm{top}} \frac{\lambda_{\pi}}{\lambda_e} \geq s_{\textrm{top}}$. 
\end{itemize}

We can see that the winner is weakly better off in the first two cases when $\lambda_{\pi} \geq s_{\textrm{top}}$ and is weakly better off in the third case whenever $\lambda_{\pi} \geq \lambda_e$. Both inequalities must hold for the Windfall Clause to make firms better off.

We solved these two inequalities to characterize the full set of rational Windfall Clauses as:
\begin{equation}
    \frac{1 - \frac{1 - e_{wc}}{e_{wc}}}{e_{wc}} \leq \tau \leq  \frac{1 - s_{\textrm{top}}}{e_{wc}} 
    \label{eq: rational set late}
\end{equation}

\begin{figure}[t]
    \centering
    \includegraphics[width=\textwidth]{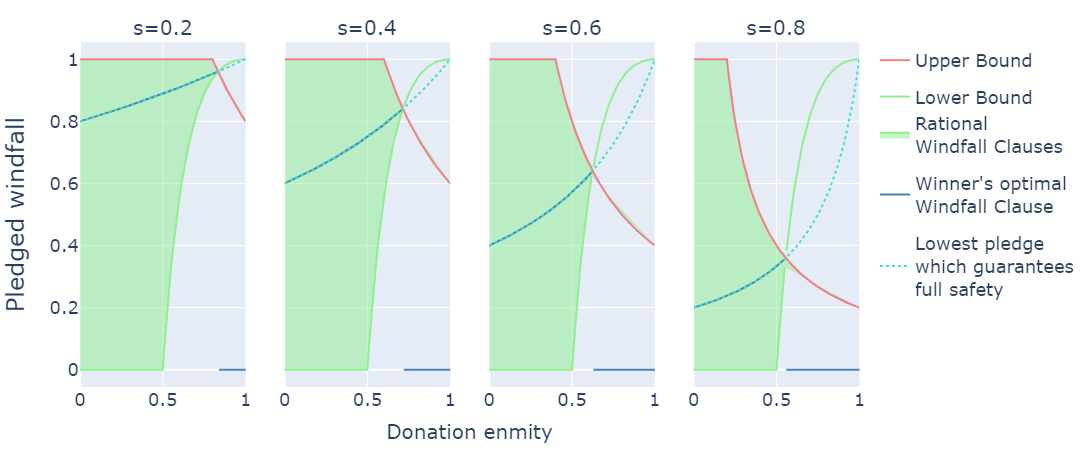}
    \caption{\small Given the winner's original safety, $s$, and the donation enmity, $e_{wc}$, a Windfall Clause is rational for all competitors as long as the pledged windfall, $\tau$, lies in the shaded region between the upper bound (red line) and lower bound (green line). This region is determined by \cref{eq: rational set late}. After the two bounds meet, no Windfall Clause is rational for the winner, so the winner would prefer no Windfall Clause ($\tau=0$). In the late Windfall Clause scenario, firms can guarantee full safety by joining a sufficiently strong Windfall Clause (dotted line which partially overlaps the blue line). In this model, the pledged windfall which maximizes the winner's utility follows this dotted line until no rational Windfall Clause exists.}
    \label{fig:rational set for late windfall clause}
\end{figure}

\cref{fig:rational set for late windfall clause} plots the set of rational Windfall Clauses, $\tau$ against $e_{wc}$ for different levels of $s_{\textrm{top}}$. In line with the previous section, \cref{sec:result2}, this set grows as $s_{\textrm{top}}$ falls.

\cref{eq: rational set late} tells us that a rational Windfall Clause exists as long as $s_{\textrm{top}} \leq \frac{1 - e_{wc}}{e_{wc}}$, or equivalently $e_{wc} \leq \frac{1}{1 + s_{\textrm{top}}}$ as shown in \cref{fig:rational set for late windfall clause}. Assuming we don't already have full safety ($s_{\textrm{top}} < 1$), a rational Windfall Clause always exists for any $e_{wc} \leq 0.5$.

We also identify the rational Windfall Clauses in \cref{fig:rational set for late windfall clause} which maximize the utility of the winning firm. Surprisingly, when a rational Windfall Clause exists, the optimal clause for the firm matches the clause with the lowest pledged windfall, $\tau$, that guarantees full safety.\footnote{Recall that the winner's safety under a late Windfall Clause is $s= \frac{s_{\textrm{top}}}{\lambda_e}$, so the lowest pledged windfall that guarantees full safety, $s=1$, is $\tau = \frac{1 - s_{\textrm{top}}}{1 - e_{wc} s_{\textrm{top}}}$. Via further calculations (not shown here), we can show that whenever a rational late Windfall Clause exists, as in \cref{eq: rational set late}, that the same expression for $\tau$ maximises the winner's payoff.} As long as donation enmity is not too high, $e_{wc} \leq \frac{1}{1 + s_{\textrm{top}}}$, it is possible to find a Windfall Clause which both guarantees full safety and is also optimal for the winner.

Since we can guarantee full safety at these pledged windfalls, it makes sense to say that the late Windfall Clause has a donation enmity limit $e_{wc}^* = \frac{1}{1 + s_{\textrm{top}}}$. Immediately, we know that the donation enmity limit must satisfy: $e_{wc}^* \geq 0.5$, which is already greater than the donation enmity limit for the early Windfall Clause in \cref{fig: expected payoffs default}. 

Note that we don't need to pledge all windfall profits, $\tau=1$, to guarantee full safety, as we did for early Windfall Clauses. Therefore, we are likely to find Windfall Clauses which make firms better off and guarantee full safety for higher values of donation enmity, increasing the donation enmity limit. Moreover, for early Windfall Clauses, any clause risks being too strong: a weaker late Windfall Clause would have led to full safety in the competition whilst sacrificing a smaller share of profits. We now have a good basis for understanding why, when firms know their capabilities, it is often far easier to find a Windfall Clause which is strong enough to encourage the winner to guarantee full safety.

\begin{figure}
    \centering
    \includegraphics[width=\textwidth]{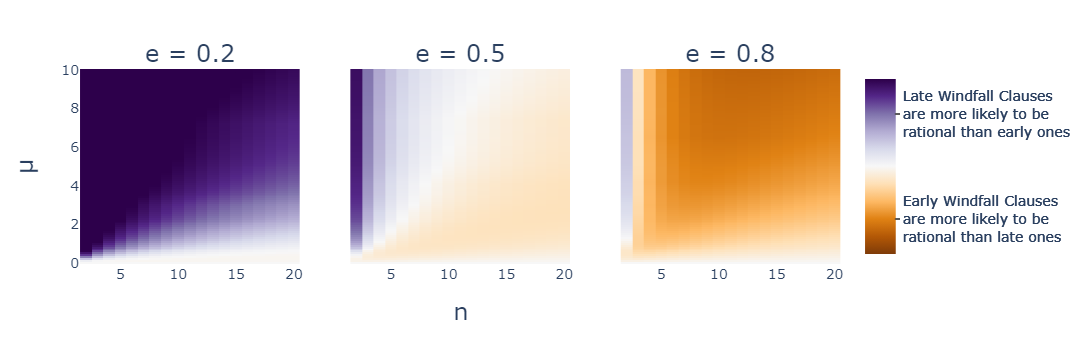}
    \caption{\small The difference between the donation enmity limits for late and early Windfall Clauses as we vary the number of competitors, $n$, and max capability, $\mu$, for different values of $e$. When $e$ is low, the donation enmity limit is often much higher for the late Windfall Clause. When $e$ is high, the donation enmity limit is somewhat higher for the early Windfall Clause, especially for high $n$. These patterns reflect that the donation enmity limit remains high for the late Windfall Clause throughout the parameter space, whereas the early Windfall Clause only has an especially high donation enmity limit when $n$ or $e$ is high.}
    \label{fig: gap in donation enmity limits robust in e}
\end{figure}

Yet, there are regions of the parameter space where knowing everyone's capabilities makes finding a rational Windfall Clause harder. \cref{fig: gap in donation enmity limits robust in e} plots the difference between the donation enmity limits for late and early Windfall Clauses. To compare late donation enmity limits to the early donation enmity limits from \cref{eq: donation enmity limit}, we first average over the late donation enmity limit, $e_{wc}^* = \frac{1}{1 + s_{\textrm{top}}} = \frac{1}{1 + \min({1, \frac{\Delta}{e})}}$, with respect to $\Delta$, the capability gap between the leader and their closest competitor. Although our main claim holds for an extensive region of the parameter space, for high $e$, and for many values of $n$ we see a smaller donation enmity limit for late compared to early Windfall Clauses. 

Nevertheless, we stand by our original claim: the advantages of early and late Windfall Clauses are not symmetric. Firms often have a stronger incentive to join a late Windfall Clause than an early one. Notice that when late Windfall Clauses have a lower donation enmity limit than early ones, the competition is very dangerous, so both donation enmity limits are already very high. On the other hand, when the average late Windfall Clause has a higher donation enmity limit than an early one, it is often much higher. Overall, it often will be easier to negotiate a late Windfall Clause than an early one (especially if $e$, $n$, or $\mu$ are uncertain).

Given then that it is usually easier to negotiate a Windfall Clause with information about capabilities than without, we may be tempted to recommend that a Windfall Clause be implemented when all information is available. However, there are many reasons not to wait until such a scenario emerges: it may take time to implement a Windfall Clause or to convince all competitors to join, and the long-term safety of TAI may require that we increase our safety R\&D as soon as possible. 

A more robust takeaway is that an early Windfall Clause is a credible commitment. Another way of framing stronger incentives to join a late Windfall Clause is that firms who had joined early would have a lot to lose from leaving that Windfall Clause. Even as firms learn about their capabilities, they would prefer to keep their current Windfall Clause than to have no such clause. Therefore, it is very plausible that if we can find an early Windfall Clause with a donation enmity low enough, firms will stick with the Windfall Clause, even as they learn their capabilities. 
In other words, they have no incentive to leave a Windfall Clause in the middle of the competition.

Note that the winning firm will desire weakening a strong early Windfall Clause as they learn everyone's capabilities, but they still prefer the original clause to having no clause at all. Yet, the ease of negotiating Windfall Clauses when we have more information implies that there may be gains from renegotiating a Windfall Clause. Extending the model to allow dynamic renegotiation could hold important lessons for making a Windfall Clause even more credible \cite{kranz_discounted_2012,yeung_subgame_2016,goldlucke_reconciliating_2020,li_multilateral_2020}.

\section{Conclusion}\label[section]{sec:conclusion}

It may surprise firms to hear that it is often in their best interests to pledge a large share of windfall profits to socially good causes. In our model, a Windfall Clause encourages a safer competition for TAI and aligns the goals of other firms with their own.

Our analysis combined a simple model of the Windfall Clause proposed by \citeA{okeefe_windfall_2020} with \citeS{armstrong_racing_2016} model of TAI competition. We found that as the competition becomes more dangerous, due for example to an increasing number of entrants or the curse of more information from \citeA{armstrong_racing_2016}, firms have an even stronger incentive to join the Windfall Clause. Moreover, as firms learn about their relative positions in the competition, they often become more willing to commit to a Windfall Clause strong enough to greatly improve safety. The Windfall Clause is therefore a highly credible policy: firms have no incentive to leave the Windfall Clause in the middle of the competition, although they may wish to renegotiate a better one (if they have the time). Finally, it can be in the best interests of firms to join the Windfall Clause even if they dislike donating via the Windfall Clause more than letting another firm win the competition. The scope for agreeing to a mutually beneficial Windfall Clause is surprisingly large.

These findings provide evidence in favor of working quickly to prepare a Windfall Clause as part of our policy toolkits, especially in case TAI competition emerges soon. Even in the most dangerous scenarios, we have a good chance of designing a Windfall Clause which firms will rationally join.

Future research on modeling a Windfall Clause which we believe will help this effort include: extending the competition model to allow for multiple winners, allowing multiple rounds to negotiate a Windfall Clause, and investigating how to implement a Windfall Clause across borders with competitors other than firms. We also encourage efforts to model the many nuances of the Windfall Clause contribution function outlined by \citeA{okeefe_windfall_2020} and efforts to design Windfall Clauses for when firms already have private information about their capabilities.

\newpage

\section*{Acknowledgments} 
We thank the team at Convergence Analysis for their assistance as a fiscal sponsor and receiving charity. Our thanks also go to Vasily Kuznetsov and Miles Tidmarsh for the time they dedicated to moving forward the research of Modeling Cooperation. We also thank Adrian Hutter, David Kristoffersson, Lewis Bova, Misha Yagudin, Tanja Rüegg, and our anonymous reviewers for their comments and suggestions. This research was funded with a grant from Jaan Tallinn as recommended by the Survival and Flourishing Fund. All mistakes in this article are our own.

\vskip 0.2in
\bibliography{library}
\bibliographystyle{theapa}

\end{document}